\documentstyle[epsf,aaspp4]{article}

\newcommand{\simgt}{\lower.5ex\hbox{$\; \buildrel > \over \sim \;$}}
\newcommand{\simlt}{\lower.5ex\hbox{$\; \buildrel < \over \sim \;$}}
\newcommand{\citet}[1] {\cite{#1}}
\newcommand{\citep}[1] {(\cite{#1})}
 \newcommand{\bm}[1]{\mbox{\boldmath$#1$}}
 \newcommand{\kaco}[1]{\left\langle{#1}\right\rangle}
 \newcommand{\skaco}[1]{\langle{#1}\rangle}
\begin{document}
\title{Lensing-induced Non-Gaussian Signatures 
in the Cosmic Microwave Background}


\author{Masahiro Takada}
\affil{
Division of Theoretical Astrophysics,
National Astronomical Observatory,
2-21-1 Osawa, Mitaka, Tokyo 181-8588, Japan}
\affil{mtakada@th.nao.ac.jp}
%
\begin{abstract}
We propose a new method for extracting the non-Gaussian signatures on the
isotemperature statistics in the cosmic microwave background
(CMB) sky, which is induced by the gravitational lensing due to
the intervening large-scale structure of the universe.
To develop the method, we focus on 
a specific statistical property of the intrinsic Gaussian CMB field;
a field point in the map that has 
a larger absolute value of the temperature threshold
tends to have a larger absolute value of the 
curvature parameter defined by a trace of second 
derivative matrix of the temperature field, while the ellipticity 
parameter similarly defined is uniformly distributed independently of 
the threshold because of the isotropic nature of the Gaussian field.
The weak lensing then causes a stronger
distortion effect on the isotemperature contours with higher
threshold and especially induces a coherent 
distribution of the ellipticity parameter 
correlated with the threshold
as a result of the coupling between the CMB curvature parameter
and the gravitational tidal shear in the observed map.
These characteristic patterns can be statistically picked up 
by considering
three independent characteristic functions, which are
 obtained from the averages
of quadratic combinations of the second derivative fields of CMB 
over isotemperature contours with each threshold. Consequently, we 
find that the lensing effect generates non-Gaussian signatures 
on those functions that have a distinct functional dependence of
the threshold. 
We test the method using numerical simulations of CMB maps
and show that the lensing signals can be measured definitely, 
provided that we use CMB data with sufficiently low noise and high
angular resolution.
\end{abstract}
%
%
\section{Introduction}
Determination of the power spectrum of dark matter fluctuations 
in the observed hierarchical large-scale structures of the 
universe remains
perhaps  the compelling problem in cosmology. Weak gravitational
lensing  due to the large-scale structure
is recognized as a powerful probe of solving this
problem as well as of constraining the cosmological parameters
(\cite{Gunn}; \cite{Blandford}; \cite{Miralda}; \cite{Kaiser92}),
because it can fully avoid uncertainties associated with 
the biasing problem. 
Recently, several independent groups have reported significant detections
of coherent gravitational distortions of distant galactic images
(\cite{Waerbeke}; \cite{Wittman}; \cite{Bacon}; \cite{Kaiser}; \cite{Maoli}).
On the other hand, the temperature anisotropies in the cosmic microwave
background (CMB) can be the most powerful probe of our universe,
especially of fundamental cosmological parameters (e.g., \cite{hunat}).
The weak lensing similarly induces distortions in
the pattern of the CMB anisotropies, and  the lensing signatures
will provide a wealth of information on inhomogeneous matter distribution
and evolutionary history of dark matter fluctuations
between the last scattering surface and present. We then expect that 
the cosmological implications provided from the measurements of lensing
effects on the CMB will be very precise, because
there is no ambiguity in theoretical understanding of the primary
CMB physics and about the distance of the source plane. 
However, it is concluded that the weak lensing effects on the CMB angular 
power spectrum $C_l$ is small (e.g. see \cite{Seljak96} and references 
therein), although 
the detailed CMB analyses need to also take into account the lensing 
contribution. 
Recently, Seljak \& Zaldarriaga (1999) (see also \cite{ZS99}) 
developed a new method for a direct
reconstruction of the projected matter power spectrum from the observed 
CMB map, and showed that it could be successfully achieved if there is 
no sufficient small scale power of intrinsic CMB anisotropies.
In this method, the lensing signals can be extracted
 by averaging quadratic combinations 
of the CMB derivative fields over many independent CMB patches like 
the analysis to extract the distortion effect on distant galactic images,
even though the reconstruction maps have a low signal to noise ratio
on individual patches. 

Excitingly, the high-precision data from the BOOMERanG (\cite{boom};
\cite{Lange})
and MAXIMA-1 (\cite{Maxima}; \cite{Balbi}) have revealed that
the measured angular power spectrum $C_l$ is fairly consistent
with that predicted by the inflation-motivated
adiabatic cold dark matter models (also see \cite{Tegmark00};
\cite{Hu00}). The standard inflationary scenarios also
predict that the primordial fluctuations are homogeneous and isotropic
Gaussian (\cite{Guth}), and then statistical properties of any CMB 
fields can be
completely determined from the two-point correlation function $C(\theta)$
or equivalently $C_l$ based on the Gaussian 
random theory (\cite{BBKS} hereafter BBKS; \cite{BE} hereafter BE).
Taking advantage of this predictability, 
various statistical methods to extract the non-Gaussian signatures
induced by the weak lensing have been proposed. Bernardeau (1998) 
found that the lensing alters a specific shape of the probability 
distribution function (PDF) of ellipticity parameter 
for field point or peak for the Gaussian case
as a result of an excess of elongated structures in the observed (lensed) map.
Although the method could be a powerful probe to measure the matter
fluctuations around the characteristic curvature scale of CMB,
the beam smearing effect of a telescope is crucial for the detection
because it again tends to circularize the deformed structures.
Van Waerbeke, Bernardeau \& Benabed (2000)
then investigated that a statistically correlated alignment between 
the CMB and distant galactic ellipticities could be detected with a 
higher signal to noise ratio, provided that a galaxy survey follow-up
can be done on a sufficiently large area. We have quantitatively
investigated the weak lensing effect on the two-point correlation function
of local maxima or minima in the CMB map, and it can potentially probe
the lensing signatures on large angular scales such as $\theta\approx 
70'$ that
corresponds to the matter fluctuations with wavelength modes around
$\lambda\sim 50h^{-1}{\rm Mpc}$ 
(\cite{TKF}; \cite{TF}). Recently, using numerical simulations,
it was shown that the lensing effect
causes a change of normalization factors for three morphological
descriptions of the CMB map, the so-called Minkowski functionals, 
against their Gaussian predictions (\cite{STF}). 

The purpose of this paper is to develop a new simple method for extracting
the lensing-induced non-Gaussian signatures from the CMB map based on 
the isotemperature
statistics.  We then focus on specific 
 statistical properties of the intrinsic Gaussian
CMB field;  a field point that has a larger absolute value of  
the temperature threshold tends to have a larger absolute 
value of the curvature parameter  defined by a trace of the 
second derivative matrix of the CMB field, while the ellipticity 
parameter similarly defined is uniformly distributed independently of 
the threshold because of the isotropic nature of the Gaussian field.
From these features, we expect that the weak lensing causes a larger
distortion effect on structures of temperature fluctuations 
around a point with higher threshold. In particular,  
the lensing can induce a coherent distribution of 
the ellipticity parameter correlated with 
the threshold owing to the coupling
between the CMB curvature and the gravitational tidal shear.
To extract these characteristic patterns, we define three 
independent functions based on the isotemperature statistics that are
obtained from the averages of quadratic combinations of
the second derivatives of CMB field over isotemperature contours
with each threshold. As a result, we find the lensing effect on those 
characteristic functions generates a definite functional dependence of 
the threshold, and therefore the lensing signals could be easily 
measured as a non-Gaussian signature since those functions have  very 
specific shapes in the Gaussian (unlensed) case.
Using numerical simulations of lensed and unlensed CMB maps
including the instrumental effects of beam smearing and detector noise,
we investigate the feasibility of the method.

This paper is organized as follows. In \S\ref{method} we 
formulate a method for extracting the lensing-induced non-Gaussian signatures
using the Gaussian random theory for the primary CMB. In \S\ref{models}
we outline the procedure of numerical experiments of our method
using the simulated CMB maps with and without the weak lensing effect. 
In \S\ref{result} we present some results in the flat universe with 
a cosmological constant and investigate the detectability
of lensing signatures by taking into account the measurement
errors associated with the cosmic variance and  the instrumental effects 
especially for the future satellite mission 
{\em Planck Surveyor}\footnote{{\tt
http://astro.estec.esa.nl/SA-general/Projects/Planck/}}. 
In the final section some discussions and conclusions are presented.

\section{Method: Weak Lensing Effect on Isotemperature Statistics}
\label{method}
\subsection{Random Gaussian Theory}
In this section, we briefly review a relevant part of the 
Gaussian random
theory developed by BBKS and BE for three- and two-dimensional cases,
respectively. First, we define the temperature fluctuation field
in the CMB map as $\Delta(\bm{\theta})\equiv [T(\bm{\theta})
-T_{\rm CMB}]/T_{\rm CMB}$. Throughout this paper we employ the flat sky
approximation developed by BE, and this is a good approximation
for our study because the lensing deformation effect on
the CMB anisotropies is important only on arcminite scales. The Fourier
transformation can be then expressed as $\Delta(\bm{\theta})\equiv\int
d^2\bm{l}/(2\pi)^2\Delta(\bm{l})e^{i\bm{l}\cdot\bm{\theta}}$,
and  the statistical properties of the unlensed CMB are completely
specified by the angular power spectrum $C_l$ defined by
$\skaco{\Delta(\bm{l})
\Delta(\bm{l}')}=(2\pi)^2 C_l\delta^2(\bm{l}-\bm{l}')$.

\begin{figure}[t]
 \begin{center}
     \leavevmode\epsfxsize=8cm \epsfbox{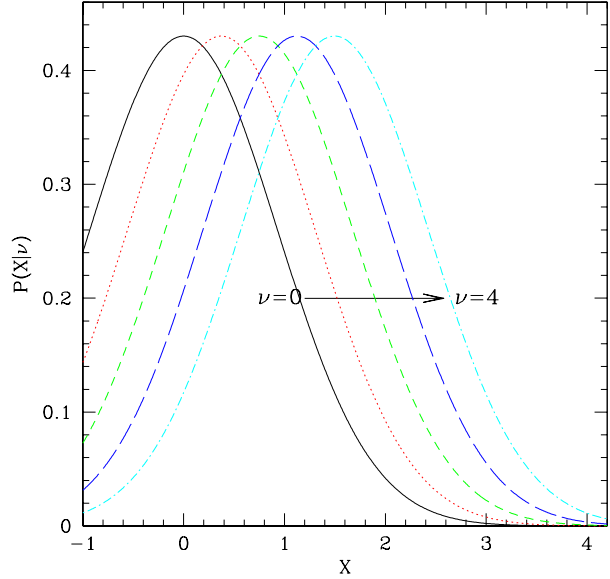}
\caption{
  The conditional probability distributions of the curvature parameter 
  $X$ for a field point with the height $\nu=0$ (solid), $\nu=1$ (dot), 
  $\nu=2$ (short dash), $\nu=3$ (long dash), and $\nu=4$ (dot-dash) 
  are plotted for $\gamma_\ast=0.35$. 
  \label{fig:peakshape}}
 \end{center}
\end{figure}

According to the Gaussian random theory,
a certain set of variables $v_{i}(i=1,2,...,N)$ constructed from 
the CMB field obeys the following joint probability distribution
function (PDF); 
\begin{equation}
p(\bm{v})=\frac{1}{(2\pi)^{N/2}|{\rm det}(M_{ij})|}
 \exp\left[-\frac{1}{2}v_iM^{-1}_{ij}v_j\right],
\label{eqn:gausstheo}
\end{equation}
where $M_{ij}$ is the covariance matrix defined by
$M_{ij}\equiv\kaco{(v_i-\kaco{v_i})(v_j-\kaco{v_j})}$, and $M^{-1}$ and
${\rm det}(M_{ij})$ denote the inverse and determinant, respectively.
Since we are interested in the lensing distortion effect on the
isotemperature contours as a function of the temperature threshold,
we pay special attention to statistical properties of the 
second derivative field of $\Delta$, because
the local curvature of CMB is probably a good indicator of the
lensing distortion effect as shown latter. 
It is then convenient to introduce the following variables
\begin{eqnarray}
&&\nu\equiv\frac{\Delta}{\sigma_0},\hspace{2em}
X\equiv-\frac{\Delta_{11}+\Delta_{22}}{\sigma_2},\hspace{2em}
Y\equiv\frac{\Delta_{11}-\Delta_{22}}{\sigma_2},\hspace{2em}
Z\equiv\frac{2\Delta_{12}}{\sigma_2},
\label{eqn:varis}
\end{eqnarray}
where $\sigma_{\rm n}$ is defined by $\sigma_{\rm
n}^2\equiv\int(ldl/2\pi)C_l l^{2{\rm n}}$, $\Delta_{ij}\equiv
\partial^2\Delta/\partial\theta^i\partial\theta^j$ and $\nu$ is 
the so-called {\em threshold} of temperature fluctuations.
To clarify the physical meanings of $X$, $Y$ and $Z$ more explicitly, 
we express them in terms of two eigenvalues 
$\lambda_1$ and $\lambda_2$ for normalized curvature matrix
$-\Delta_{ij}/\sigma_2$ as
\begin{equation}
X=-(\lambda_1+\lambda_2),\hspace{2em}Y=-2eX\cos(2\varphi),\hspace{2em}
 Z=-2eX\sin(2\varphi),
\end{equation}
where $e$ represents the local ellipticity parameter defined by $e\equiv
(\lambda_1-\lambda_2)/[2(\lambda_1+\lambda_2)]$ and $\varphi$ denotes
the relative angle between the principal axis of $\Delta_{ij}$ and the $1$-axis.
By the meaning of equation above, hereafter we call $X$ 
a local {\em curvature parameter} around a given field point. 
Moreover, if the isotemperature contour in the neighborhood of local
minima or maxima is given
by an ellipse of $f(\theta_1,\theta_2)=\theta_1^2/b^2+\theta_2^2/a^2$ in 
the coordinates of principal axes, the parameter $e$ can be expressed
in terms of $a$ and $b$ as $e=(a^2-b^2)/[2(a^2+b^2)]$.
Hence, $Y$ and $Z$ represent $1$- and
$2$-axis components of the local ellipticity parameter
of temperature curvature field, respectively.  
The non-zero second moments of the variables (\ref{eqn:varis}) can be 
then calculated as
\begin{equation}
\kaco{\nu^2}=\kaco{X^2}=2\kaco{Y^2}=2\kaco{Z^2}=1,\hspace{2em}
 \kaco{\nu X}=\gamma_\ast,
\label{eqn:secondmom} 
\end{equation}
where $\gamma_\ast=\sigma_1^2/(\sigma_0\sigma_2)$ (although we will also 
use the same letter $\gamma$ for a shear component of lensing deformation
tensor, we want readers not to confuse $\gamma_\ast$ and $\gamma$) and
$\gamma_\ast$ represents the strength of cross 
correlation between $X$ and $\nu$
from the relation of $\gamma_\ast=\skaco{\nu X}$.
Equation (\ref{eqn:gausstheo}) tells us that 
the joint PDF of variables $v_i=(\nu,X,Y,Z)$ for one field
point becomes
\begin{equation}
p(\nu,X,Y,Z)=\frac{2}{(2\pi)^2\sqrt{1-\gamma_\ast^2}}\exp[-Q],
\label{eqn:pdf}
\end{equation}
with
\begin{equation}
2Q\equiv \nu^2+\frac{(X-\gamma_\ast\nu)^2}{(1-\gamma_\ast^2)^2}+2Y^2+2Z^2.
\end{equation}
The important result is that $\nu$ and $X$ have the
non-vanishing cross correlation, and the term of
$\exp[-(X-\gamma_\ast\nu)^2/(2(1-\gamma_\ast^2))]$ in
equation (\ref{eqn:pdf}) physically means that structures around a
field point with larger absolute threshold 
tend to have a larger absolute value of the curvature parameter $X$. 
In fact, this feature is more explicitly clarified by considering
the conditional probability distribution for field points with
 a given threshold $\nu$.
Figure \ref{fig:peakshape} shows the distribution
of curvature parameter $X$ subject to the constraint that the point
has a given threshold $\nu$, where the conditional PDF is defined by
$p(X|\nu)\equiv p(\nu,X)/p(\nu)=1/\sqrt{2\pi(1-\gamma_\ast^2)}
\exp[-(X-\gamma_\ast\nu)^2/(2(1-\gamma_\ast^2))]$.
The absence of correlation between $\nu$ and $Y$ or $Z$ is the consequence of
the isotropic nature of Gaussian field, more specifically due to
the isotropic distribution of an orientation angle of ellipticity
parameter. 

Using the PDF (\ref{eqn:pdf}), we define the
following three independent functions with respect to
temperature threshold $\nu_{\rm
t}$ that characterize statistical properties of second derivative
fields of CMB along isotemperature contours with the threshold $\nu_{\rm
t}$;
\begin{eqnarray}
&&V_{X}(\nu_{\rm t})=\kaco{\delta(\nu-\nu_{\rm t})X^2}=
 \frac{1}{\sqrt{2\pi}}\exp\left(
-\frac{\nu_{\rm t}^2}{2}\right)\left[(1-\gamma_\ast^2)
+\gamma_\ast^2\nu_{\rm t}^2 \right]\nonumber \\
&&V_{Y}(\nu_{\rm t})=\kaco{\delta(\nu-\nu_{\rm t})Y^2}=
 \frac{1}{2\sqrt{2\pi}}\exp\left(
-\frac{\nu_{\rm t}^2}{2}\right)\nonumber\\
&&V_{Z}(\nu_{\rm t})=\kaco{\delta(\nu-\nu_{\rm t})Z^2}=V_Y(\nu_{\rm t}),
\label{eqn:isofun}
\end{eqnarray}
where the bracket is defined by $\skaco{\cdots}\equiv \int\!\!d\nu dXdYdZ
\cdots p(\nu,X,Y,Z)$ and can be observationally interpreted as 
average of the considered local quantities 
performed over all the isocontours in the CMB sky 
from the assumption of large scale statistical homogeneity. 
All other averages of 
quadratic combinations of the second derivatives such as
$\kaco{XY}=\kaco{XZ}=\kaco{YZ}$ vanish because of the isotropic nature of
the Gaussian field.  The functions in equation (\ref{eqn:isofun})
thus have very specific shapes for the Gaussian case, and 
we can take advantage of this property in order to extract the
non-Gaussian signatures on those functions
 induced by the lensing distortion effect.

\subsection{Lensing distortion effect on the isotemperature contours as a
  non-Gaussian signature}

The CMB photon rays are randomly deflected by
the inhomogeneous matter distributions inherent in the
intervening large-scale structures of the universe during their propagations 
from the last scattering surface to us. 
Therefore, the observed CMB temperature fluctuation field at a certain
angular direction $\bm{\theta}$, $\tilde{\Delta}(\bm{\theta})$, is equal to
the primary field emitted from the another direction $\bm{\theta}
+\bm{\xi}(\bm{\theta})$ on the last scattering surface,
$\Delta(\bm{\theta}+\bm{\xi})$, where $\bm{\xi}(\bm{\theta})$ is
the displacement field. The lensed second derivative
field of CMB can be then expressed by
\begin{eqnarray}
\tilde{\Delta}_{ij}
&=&(\delta_{im}+\xi_{m,i})\Delta_{mn}(\delta_{nj}+\xi_{n,j})
 +\Delta_{m}\xi_{m,ij}\nonumber\\
&=&{\cal A}_{im}\Delta_{mn}{\cal A}_{nj}
+\Delta_{m}\xi_{m,ij},
\label{eqn:glpr}
\end{eqnarray}
where ${\cal A}$ is the so-called amplification matrix and
$\delta_{ij}$ is the Kronecker delta symbol. Hereafter, the variables
with and without tilde symbol denote the lensed and unlensed CMB fields, 
respectively. 
The components of ${\cal A}_{ij}$ can be expressed in terms of 
the local gravitational convergence $\kappa$ and tidal distortion
$\gamma$ as
\begin{equation}
{\cal A}_{ij}=
 \left(
\begin{array}{cc}
 1-\kappa-\gamma_1&-\gamma_2\\
 -\gamma_2&1-\kappa+\gamma_1
  \end{array}
 \right). 
\end{equation}
In the weak lensing regime the matrix ${\cal A}$ is always
regular, and the variances of $\kappa$, $\gamma_1$ and $\gamma_2$ are
related to each other through
\begin{equation}
\kaco{\kappa^2}=2\kaco{\gamma_1^2}=2\kaco{\gamma_2^2}=\sigma^2_\kappa.
\end{equation}
We here have not assumed that $\kappa$ and $\gamma_i$ are Gaussian, and
this is a simple consequence of statistical isotropy of the displacement
field. As shown by the several works using the ray tracing simulations 
through the large-scale structure modeled by N-body simulations, 
the lensing fields are indeed not Gaussian on angular scales
of $\theta\simlt 10'$ (\cite{JSW}; \cite{Hamana}), and 
 we will later discuss the problem how the non-Gaussian features of $\kappa$
could affect our results.
The second moment of the convergence, $\sigma_\kappa$, is related to
the projected matter power spectrum (\cite{Kaiser92}):
\begin{equation}
\sigma_\kappa^2=\int\!\!\frac{ldl}{2\pi}P_\kappa(l)
=\frac{9}{4}H_0^4\Omega_{m}^2\int\!\!\frac{ldl}{2\pi}\int^{\chi_{\rm rec}}
_0\!\!d\chi a^{-2}(\tau)W^2(\chi,\chi_{\rm rec})P_\delta\!\left(k
=\frac{l}{r(\chi)},\chi\right),
\label{eqn:conv}
\end{equation}
where $P_\delta(k)$ and $P_{\kappa}(l)$ denote the three-dimensional power
spectrum of matter fluctuations and its projected power spectrum,
respectively. $\tau$ is a conformal time, $\chi\equiv\tau_0-\tau$,
and the subscripts $0$ and ``r'' denote values at present and the
recombination time, respectively. $H_0(=100h{\rm km~s}^{-1}{\rm
~Mpc}^{-1})$ and $\Omega_{\rm m0}$ denote the present-day Hubble constant
and energy density parameter of matter, respectively.  $r(\chi)$ is the
corresponding comoving angular diameter distance, defined as
$K^{-1/2}\sin K^{1/2}\chi$, $\chi$, $(-K)^{-1/2}\sinh(-K)^{1/2}\chi$
for  $K>0$, $K=0$, $K<0$, respectively, where the curvature parameter
$K$ is represented as $K=(\Omega_{m0}+\Omega_{\lambda0}-1)H_0^2$ and
$\Omega_{\lambda0}$ is the present-day vacuum energy density relative to 
the critical density. 
The projection operator $W(\chi,\chi_{\rm rec})$ on the celestial
sphere is given by $W(\chi,\chi_{\rm rec})=r(\chi_{\rm
rec}-\chi)/r(\chi_{\rm rec})$. As shown later, the effect of the finite
beam size $\theta_{\rm fwhm}$ of a telescope on $\sigma_\kappa$
appears as a cutoff at $l\simgt l_{\rm sm}$ in the integration of
equation (\ref{eqn:conv}) from  the relation of $l_{\rm sm}\sim 1/\theta_{\rm
fwhm}$  and thus
$\sigma_{\rm \kappa}$ also depends on $\theta_{\rm fwhm}$ in a
general case. Inversely, by changing the smoothing scale artificially,
we could reconstruct the scale dependence of projected matter power
spectrum and we will also investigate this possibility.
The important result of equation (\ref{eqn:conv}) is that the 
magnitude of $\sigma_\kappa$ is 
sensitive to $\Omega_{m0}$ and particularly to the normalization of matter power
spectrum of $P_\delta$, which is conventionally expressed in terms of
the rms mass fluctuations of a sphere of $8h^{-1}{\rm Mpc}$, i.e.,
$\sigma_8$. Similarly, variances of the second derivative fields of
displacement field $\xi_i$ can be calculated as
\begin{equation}
 \kaco{\xi_{1,11}^2}=5\kaco{\xi_{1,12}^2}=5\kaco{\xi_{1,11}\xi_{1,22}}
  =5\kaco{\xi_{1,12}\xi_{2,22}}=\frac{5}{16}s^2
 \end{equation}
with
\begin{equation}
s^2\equiv4\int\frac{ldl}{2\pi}l^4P_\kappa(l).
\end{equation}

Equation (\ref{eqn:glpr}) yields the following relations between
the lensed (observed) and primary components of the second curvature
matrix of temperature fluctuations up to the second order of $\xi$:
\begin{eqnarray}
&&\tilde{X}=\left[(1-\kappa)^2+\gamma^2\right]X+2\gamma_1 Y
 +2\gamma_2 Z
 -\frac{\Delta_{,i}}{\sigma_2}\left(\xi_{i,11}+\xi_{i,22}\right),\nonumber\\
&&\tilde{Y}=\left[(1-\kappa)^2+\gamma_1^2-\gamma_2^2\right]Y
 +2\gamma_1 X+\frac{\Delta_{,i}}{\sigma_2}\left(\xi_{i,11}-\xi_{i,22}
\right),
 \nonumber\\
&&\tilde{Z}=\left[(1-\kappa)^2-\gamma_1^2+\gamma_2^2\right]Z
 +2\gamma_2 X+\frac{\Delta_{,i}}{\sigma_2}\xi_{i,12}.
\label{eqn:lens2deri}
\end{eqnarray}
where we have ignored the second order contributions of
$\kappa\gamma_i$, $\gamma_1\gamma_2$ and so on because they vanish after
the average as a consequence of the statistical isotropy of $\xi_i$.
Note that the weak lensing 
does not change the relations between second moments of these lensed 
variables compared with the Gaussian cases (\ref{eqn:secondmom});
$\skaco{\tilde{X}^2}=2\skaco{\tilde{Y}^2}=2\skaco{\tilde{Z}^2}\approx
1+10\sigma_\kappa^2$. The equation (\ref{eqn:lens2deri}) for 
$\tilde{Y}$ or $\tilde{Z}$ implies that the lensing effect could 
induce an ellipticity parameter at a certain field point that
arises from a coupling between the curvature parameter $X$ 
and the gravitational shear $\gamma$ even if the intrinsic
ellipticity is zero ($Y=Z=0$). Since
this effect is observable only in a statistical sense, we focus
our investigations on the problem how the lensing alters statistical 
properties of the CMB second derivative fields based on 
the isotemperature statistics.

Now we present theoretical predictions of lensed 
functions (\ref{eqn:isofun}) with respect to temperature threshold 
in the observed CMB map. If we 
assume that the primary CMB fields on the last 
scattering surface and the lensing displacement field due to the 
large-scale structure are statistically independent, 
after straightforward calculations we can obtain
\begin{eqnarray}
&&\tilde{V}_{X}(\nu_{\rm t})=\kaco{\delta(\tilde{\nu}
-\nu_{\rm t})\tilde{X}^2}=\frac{1}{\sqrt{2\pi}}
\exp\left(-\frac{\nu_{\rm t}^2}{2}
\right)\left[(1+8\sigma_\kappa^2)\left\{(1-\gamma_\ast^2)
+\gamma_\ast^2\nu_{\rm t}^2\right\}+2\sigma_\kappa^2
+\frac{\sigma_1^2}{2\sigma_2^2}s^2\right],\nonumber\\
&& \tilde{V}_{Y}(\nu_{\rm t})=
 \kaco{\delta(\tilde{\nu}-\nu_{\rm t})\tilde{Y}^2}=
 \frac{1}{2\sqrt{2\pi}}\exp\left(-\frac{\nu_{\rm t}^2}{2}
\right)\left[(1+6\sigma_\kappa^2)+4\sigma_\kappa^2
\left\{(1-\gamma_\ast^2)+\gamma_\ast^2\nu_{\rm t}^2\right\}
+\frac{\sigma_1^2}{4\sigma_2^2}s^2\right]
 ,\nonumber\\
&&\tilde{V}_{Z}(\nu_{\rm t})=\tilde{V}_{Y}(\nu_{\rm t}).
\label{eqn:lensisofun}
\end{eqnarray}
We so far have used the perturbations only for the lensing displacement
field $\xi$ and thus these equations (\ref{eqn:lensisofun}) are valid for an
arbitrary threshold $\nu_{\rm t}$. Equation (\ref{eqn:lensisofun})
clearly shows that one of the lensing effects on these functions is the change of 
their normalization factors. The another important effect is that the
lensing generates a characteristic functional dependence of $\nu_{\rm t }$ 
on $V_X(\nu_{\rm t})$, $V_{Y}(\nu_{\rm t})$ and $V_{Z}(\nu_{\rm t})$.
This is 
as a consequence of the coupling between the CMB curvature $X$ and 
the gravitational tidal shear $\gamma$ through the intrinsic
correlation between $\nu_{\rm t}$ and $X$, 
and physically means that the lensing 
effect distorts more strongly the isotemperature contours that have larger
absolute threshold.

In practice it will be rather difficult to discriminate the change
of normalization factors caused by the lensing in equation
(\ref{eqn:lensisofun}) from the Gaussian case, because measurements of the
normalizations in the CMB map might be also sensitive to the systematic
errors, for example, due to
 the discrete effect of pixel in the map. For this reason, we here focus on
the non-Gaussian signatures that have a distinct 
functional dependence of $\nu_{\rm
t}$, and consider the following observable functions normalized
 by their values at $\nu_{\rm t}=0$ as a deviation from 
the specific function $\exp[-\nu_{\rm t}^2/2]$:
\begin{eqnarray}
&&F_{X}(\nu_{\rm t})\equiv\frac{\tilde{V}_{X}(\nu_{\rm t})}
 {\tilde{V}_{X}(0)}-
\exp\left(-\frac{\nu_{\rm t}^2}{2}\right)
 \approx\exp\left(-\frac{\nu_{\rm t}^2}{2}
\right)\frac{(1+8\sigma_\kappa^2)\gamma_\ast^2
 \nu_{\rm t}^2}{
 1-\gamma_\ast^2+10\sigma_\kappa^2-8\sigma_\kappa^2\gamma^2_\ast},\nonumber\\
 &&F_{Y}(\nu_{\rm t})\equiv\frac{\tilde{V}_{Y}(\nu_{\rm t})}
 {\tilde{V}_{Y}(0)}-
\exp\left(-\frac{\nu_{\rm t}^2}{2}\right)
\approx\exp\left(-\frac{\nu_{\rm t}^2}{2}\right)
\frac{4\sigma_\kappa^2\gamma_\ast^2\nu_{\rm t}^2}{1+10\sigma_\kappa^2
-4\sigma_\kappa^2\gamma^2_\ast},\nonumber\\
&&F_{Z}(\nu_{\rm t})\equiv\frac{\tilde{V}_{Z}(\nu_{\rm t})}
 {\tilde{V}_{Z}(0)}-
\exp\left(-\frac{\nu_{\rm t}^2}{2}\right)
=F_{Y}(\nu_{\rm t}),
 \label{eqn:glfun}
\end{eqnarray}
where we have 
neglected the terms including contributions of $s^2$
in equation (\ref{eqn:lensisofun}) because we numerically confirmed that
the contributions are always small. 
In the following discussions, these three independent
functions (\ref{eqn:glfun}) are compared to the results of numerical
experiments. 
Most importantly, equation (\ref{eqn:glfun}) shows that,
although for the Gaussian case in the absence of the lensing we should have
$F_{Y}(\nu_{\rm t})=F_{Z}(\nu_{\rm t})=0$ for all $\nu_{\rm t}$ because
of $\sigma_\kappa=0$, 
the weak lensing induces distinct non-Gaussian signatures expressed as 
$\propto \nu_{\rm t}^2\gamma_\ast^2\exp[-\nu_{\rm t}^2/2]$.
Therefore, those two functions can be direct measures of the lensing
distortion effect on the isotemperature contours. 
The property of $F_Y(\nu_{\rm t})=F_Z(\nu_{\rm t})$ arises from the statistical
random orientations of both the CMB ellipticity parameter 
and the gravitational tidal
shear, and we can use the relation to distinguish or check the lensing signals
against other possible secondary non-Gaussian contributions. The lensing effects
on $F_X$, $F_Y$ and $F_Z$ depend on two parameters of $\gamma_\ast$ and
$\sigma_\kappa$. Then, note that $\gamma_\ast$ is
a {\em parameter} of the primordial CMB anisotropy field, 
which is not observable, and is related to the corresponding
 direct observable quantity $\tilde{\gamma}_\ast$ in the
lensed CMB map by
$\tilde{\gamma}_\ast\approx\gamma_\ast(1-7\sigma_\kappa^2/2)$, 
where $\tilde{\gamma}_\ast$ is defined by 
$\tilde{\gamma}_\ast=\tilde{\sigma}_1^2/(\tilde{\sigma}_0\tilde{\sigma}_2)$
from $\tilde{\sigma}_0^2=\skaco{\tilde{\Delta}^2}$, $\tilde{\sigma}_1^2
=\skaco{(\nabla\tilde{\Delta})^2}$
and $\tilde{\sigma}_2^2=\skaco{(\nabla^2\tilde{\Delta})^2}$.
We will therefore have to treat $\gamma_\ast$
as a free parameter in performing the fitting between theoretical
predictions (\ref{eqn:glfun}) and
numerical results for the functions $F_X(\nu_{\rm t})$, $F_Y(\nu_{\rm
t})$ and $F_Z(\nu_{\rm t})$
in order to determine $\sigma_\kappa$. 
We have then confirmed that $\gamma_\ast$ is well constrained mainly by
$F_X(\nu_{\rm t})$. 

\subsection{Effect of filtering}
\label{filter}
We so far  have ignored the effects of filtering.
Actual CMB temperature maps will be observed with a
finite angular resolution, or the artificial filtering method might be 
used to reduce the detector noise effect (\cite{Barreiro98}). 
The measured temperature map is then given by
\begin{equation}
\Delta^{{\cal F}}(\bm{\theta};\theta_s)
 =\int \!\!d^2\bm{\theta}'W(|\bm{\theta}-\bm{\theta}'|; \theta_{\rm s})
 \Delta(\bm{\theta}'),
\end{equation}
where $W(\theta; \theta_{\rm s})$ is a window
function and throughout this paper
we adopt the Gaussian beam approximation expressed by
$W(\theta; \theta_{\rm s})=\exp[-\theta^2/(2\theta_{\rm s}^2)]/(2\pi
\theta_{\rm s}^2)$. For the filtering of a telescope,
the smoothing angle $\theta_{\rm s}$ can be expressed in terms of
its full-width at half-maximum angle $\theta_{\rm fwhm}$ as
$\theta_s=\theta_{\rm fwhm}/\sqrt{8\ln 2}$.
The filtered lensed temperature field is given by
\begin{equation}
\tilde{\Delta}^{{\cal F}}(\bm{\theta})=\int\!\!d^2\bm{\theta}'
 W(|\bm{\theta}-\bm{\theta}'|; \theta_{\rm s})\Delta\left(\bm{\theta}'
 +\bm{\xi}(\bm{\theta}')\right).
\label{eqn:fil1}
\end{equation}
Similarly, the filtered second derivatives field of the CMB can be
expressed as
\begin{eqnarray}
\tilde{\Delta}^{{\cal F}}_{,ij}(\bm{\theta})&=&\int\!\!d^2\bm{\theta}'
 W(|\bm{\theta}-\bm{\theta}'|; \theta_{\rm s}){\cal A}_{im}(\bm{\theta}')
 \Delta_{,mn}\!\left(\bm{\theta}'+
\bm{\xi}(\bm{\theta}')\right){\cal A}_{nj}(\bm{\theta}')
 \nonumber \\
 &=&\int\!\!d^2\bm{\theta}'W(|\bm{\theta}-\bm{\theta}'|; \theta_{\rm s})
  \left[\Delta_{,ij}(\bm{\theta}')+\Delta_{,in}(\bm{\theta}')
   \xi_{n,j}(\bm{\theta}')+\cdots\right]\nonumber \\
 &=&\int\!\!\frac{d^2\bm{l}}{(2\pi)^2}W(\bm{l};\theta_{\rm s})
  \Delta_{ij}(\bm{l})e^{i\bm{l}\cdot\bm{\theta}}+\int\!\!
  \frac{d^2\bm{l}}{(2\pi)^2}\frac{d^2\bm{l}'}{(2\pi)^2}W(|\bm{l}+\bm{l}'|;
  \theta_{\rm s })\Delta_{in}(\bm{l})\xi_{n,j}(\bm{l}')e^{i(\bm{l}+\bm{l}')\cdot
  \bm{\theta}}+\cdots,
 \label{eqn:fil2}
\end{eqnarray}
where in the first line of right hand side
we have used the part integration and assumed that the surface
integral is equal to zero for a large CMB survey sky.  
These equations (\ref{eqn:fil1}) and (\ref{eqn:fil2})
mean that the filtering procedure and the lensing
effect on the CMB do not commute in a general case. Especially,
the last line in the right hand side of equation (\ref{eqn:fil2}) shows that 
the information about a certain mode $\bm{l}'$ of the lensing field
$\xi$ is coupled to sidebands of the different $\bm{l}$ modes of the CMB
field. The problem of mode coupling therefore have to be carefully
investigated for accurate measurements of our method. However, since
the intrinsic CMB anisotropies have a small
scale cutoff due to the Silk damping and the directions of the CMB
curvature and the lensing deformation field are statistically uncorrelated,
we could employ a simple approximation that the filter function $W$ in the
equation (\ref{eqn:fil2}) is applied both to the CMB intrinsic field and 
the lensing field independently.
The variance of convergence field in equations (\ref{eqn:glfun}) can be then  
expressed by
\begin{equation}
\sigma^2_{\kappa}(\theta_{\rm s})=\int\!\!\frac{d^2\bm{l}}{(2\pi)^2}
 W^2(l; \theta_{\rm s})P_\kappa(l),
\label{eqn:smconv}
\end{equation}
where $W(l; \theta_{\rm s})=\exp[-l^2\theta_{\rm s}^2/2]$.
Unfortunately, this approximation (\ref{eqn:smconv}) might not be so
accurate since the numerical experiments showed that the magnitude of 
the convergence field reconstructed by the non-Gaussian signatures
in the simulated maps
is smaller than the value of $\sigma_\kappa$ computed by equation
(\ref{eqn:smconv}).
Therefore, the validity or improvement  of this approximation should 
be further investigated using numerical experiments.

\section{Models and Numerical Experiments}
\label{models}
\subsection{Cosmological models}
To make quantitative investigations, we consider some specific cosmological
models. For this purpose, we adopt the current favored flat universe
in the adiabatic cold dark matter model, where 
the background cosmological parameters
are chosen as $\Omega_{\rm m0}=0.3$, $\Omega_{\rm \lambda 0}=0.7$,
$h=0.7$, respectively. The flat universe is strongly
supported by the recent high precision measurements of $C_l$
(\cite{boom}; \cite{Maxima}). 
The baryon density is chosen to satisfy $\Omega_{\rm b0}h^2=0.019$, which
is consistent with values obtained from the measurements of the primeval 
deuterium abundance (\cite{tytler}). To compute $C_l$ used to make
realizations of numerically simulated CMB maps,
 we used helpful CMBFAST code 
developed by Seljak \& Zaldarriaga (1996).  
As for the matter power spectrum used to compute lensing
contributions to both numerical and theoretical predictions,
we employed the Harrison-Zel'dovich spectrum and
the BBKS transfer function with the shape parameter from Sugiyama
(1995). The free parameter is only the normalization of the
present-day matter power spectrum, i.e., $\sigma_8$.
The nonlinear evolution of the power spectrum is modeled using the fitting
formula given by Peacock \& Dodds (1996).

\subsection{Numerical simulations of CMB map with and without lensing}

We perform numerical simulations of the CMB maps with and
without the lensing effect following the procedure presented
in Takada \& Futamase (2001) in detail. 
First, a realistic temperature map on a
fixed square grid can be generated from a given power spectrum, $C_l$,
based on the Gaussian assumption. Each map is initially $60\time 60~{\rm 
deg}^2$ area, with a pixel size of about $0.88~{\rm arcmin}$ $(=60~{\rm
deg}/4096)$. A two-dimensional lensing displacement 
field can be also generated as a realization of a Gaussian process using 
the power spectrum of convergence field, $P_\kappa$,
defined by equation (\ref{eqn:conv}). Note that we have now employed
a technical simplification that the displacement field is assumed to be 
Gaussian. As explained,
the lensing effect is then computed as a mapping between the observed
and primordial temperature field expressed by
$\tilde{\Delta}(\bm{\theta})=\Delta(\bm{\theta}+\bm{\xi})$. The 
temperature field on a regular grid
in the lensed map is then given by a primordial field on 
a irregular grid using a simple local linear interpolation of the
temperature field in the neighbors (so-called could-in-cell
interpolation). In the case of taking into account the instrumental
effects of beam smearing and detector noise, we furthermore smooth
out the temperature map by convolving the Gaussian window function
and then add randomly the noise field into each pixel.

\begin{figure}[t]
 \begin{center}
     \leavevmode\epsfxsize=15cm \epsfbox{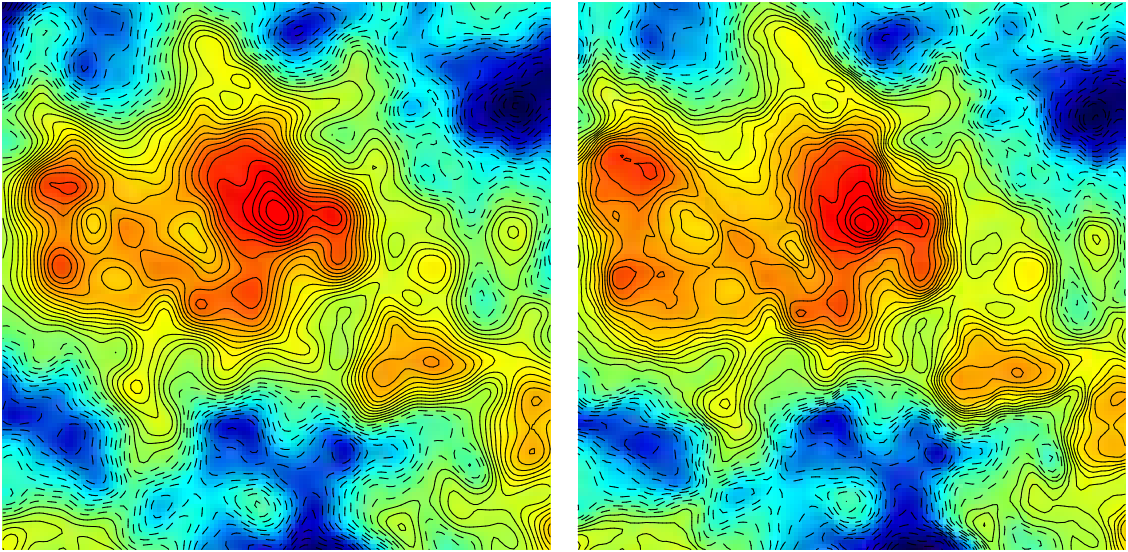}
\caption{
  An example of simulated primordial CMB anisotropies map (left) and
  the lensed map (right). We here adopted the adiabatic CDM flat universe 
  model with $\Omega_{m0}=0.3$, $\Omega_{\lambda0}=0.7$, 
  and $h=0.7$. The figures are on a side of $2$ degree and 
  the isotemperature contours are drawn in intervals of $\Delta \nu_{\rm t}
  =0.2$ $(\nu_{\rm t}\equiv\Delta(\bm{\theta})/\sigma_0)$.
  \label{fig:cmbmap}}
 \end{center}
\end{figure}

Figure \ref{fig:cmbmap} shows an example of simulated unlensed (left)
and lensed (right) CMB maps, where the isotemperature contours are also
drawn in steps of  $\Delta \nu_{\rm t}=0.2$.  This figure
illustrates that the regions around crowded contours with higher absolute
temperature threshold and larger 
curvatures are more strongly deformed by the lensing. 
Previous works (\cite{SZ00}; \cite{Zald00}) have pointed out another
but partly similar feature that the power of anisotropies on small scales 
generated by the lensing is correlated with larger scale gradient of 
the intrinsic CMB field.

\subsection{The CMB curvature field}
To calculate the second derivative fields of CMB  at a certain 
pixel in the simulated maps, we used a method of finite differences between
neighboring pixels around the point:
\begin{eqnarray}
&&\Delta_{,11}(i,j)=\left[
\Delta(i-1,j+1)-2\Delta(i,j)+\Delta(i+1,j)\right]/\delta x^2,\nonumber\\
&&\Delta_{,12}(i,j)=\frac{1}{4\delta x^2}\left[\Delta(i-1,j-1)-\Delta(i-1,j+1)
  -\Delta(i+1,j-1)+\Delta(i+1,j+1)\right],
\end{eqnarray}
where $\delta x$ is the pixel size and
$\Delta(i,j)$ is the local temperature fluctuations at the grid
point $(i,j)$. 

\section{Results and Cosmological Implications}
\label{result}
\subsection{Numerical results}
\begin{figure}[t]
 \begin{center}
     \leavevmode\epsfxsize=9.cm \epsfbox{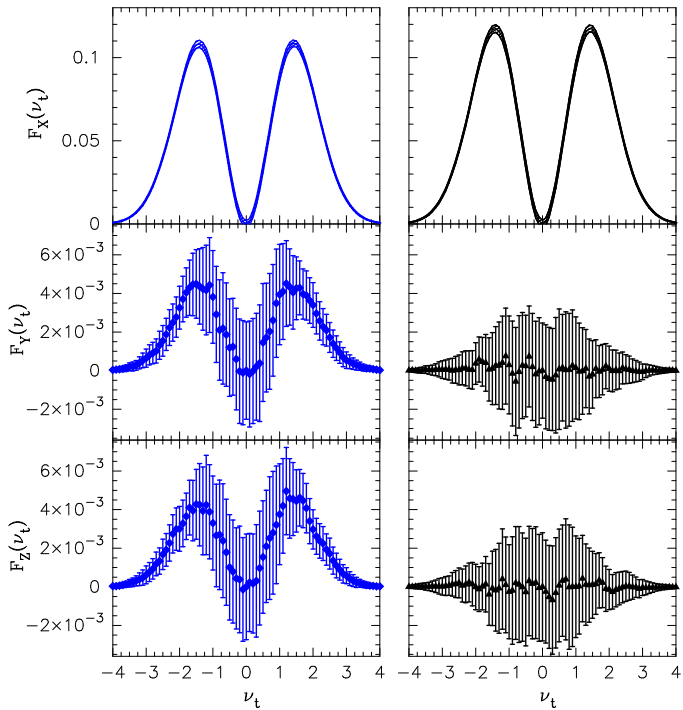}
\caption{
  The lensed (left) and unlensed (right) shapes of three functions 
  (\ref{eqn:glfun}) on the isotemperature statistics with the filter 
  scale of $3$ arcmin and $\sigma_8=2.0$ with respect to 
  the temperature threshold $\nu_{\rm t}$, which are
  computed using the numerical simulations of CMB maps. 
  The background cosmological parameters are
  $\Omega_{\rm m}=0.3$, $\Omega_{\lambda 0}=0.7$ and  $h=0.7$.
  The triangle and round marks in each panel correspond to the averages
  obtained from each $150$ independent realizations of 
  the unlensed and lensed maps, respectively.
  The error bars denote the $1\sigma$ errors due
  to  the cosmic variance computed for $70\%$ 
  sky coverage of the CMB survey. 
  The lensed curves of $F_Y(\nu_{\rm t})$ and $F_Z(\nu_{\rm t})$
  clearly show that the lensing generates a significant functional
  dependence of $\nu_{\rm t}$
  approximately expressed by the from of $\propto \nu_{\rm t}^2
  \exp[-\nu_{\rm t}^2/2]$, whereas the unlensed shapes
  have random errors with both positive and negative values around
  $F_Y(\nu_{\rm t})=F_Z(\nu_{\rm t})=0$ in each bin of $\nu_{\rm t}$.
 \label{fig:funglt3}}
 \end{center}
\end{figure}

\begin{figure}[t]
 \begin{center}
     \leavevmode\epsfxsize=8.cm \epsfbox{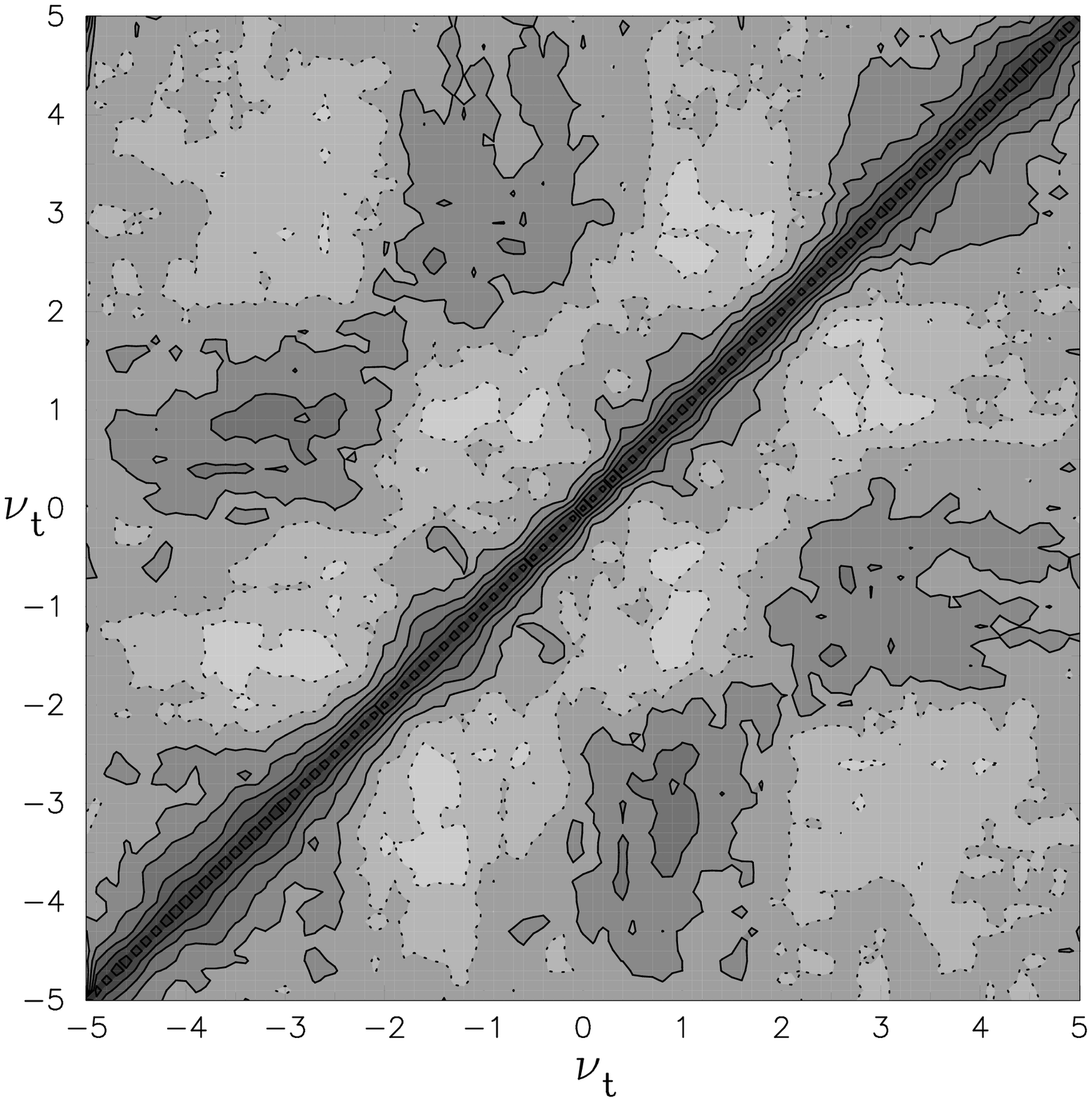}
\caption{
 Contours of correlation coefficients, $\skaco{(F_{Xi}
 -\skaco{F_{Xi}})(F_{Xj}-\skaco{F_{Xj}})}/(\sigma_i\sigma_j)$,
  for estimators of the function $F_X(\nu_{\rm t})$ defined 
  by equation (\ref{eqn:glfun}). 
  Here $F_{Xi}$ and $\sigma_i$ denote the values and variance computed
  from the simulated CMB map in the threshold bin $\nu_{\rm ti}$. The 
  contour is stepped in units of $0.2$ and the solid and dotted contours
  denote positive and negative values of the coefficients.  
 \label{fig:covmat}}
 \end{center}
\end{figure}

In Figure \ref{fig:funglt3}, we show the numerical results 
of the lensed or unlensed three functions
(\ref{eqn:glfun}) with respect to the temperature
threshold $\nu_{\rm t}$, which are obtained from each $150$ realizations of
CMB maps with $60\times 60~{\rm deg}^2$ area for the filter scale of $3$ 
arcmin. All the curves are plotted at intervals of $\Delta\nu_{\rm t}=0.1$. 
The error bar in each bin corresponds to the cosmic variance 
associated with the measurements and is estimated by
rescaling the variances obtained from all the realizations by
a factor $1/8$ when we assume the sky coverage of $70\%$ ($f_{\rm sky}=0.7$)
for a survey of the CMB sky. The figure clearly shows
that the lensing deformation effect generates a significant functional
dependence of $\nu_{\rm t}$ on $F_Y(\nu_{\rm t})$ and $F_{Z}(\nu_{\rm
t})$ approximately expressed in the form proportional to $\nu_{\rm t}^2
\exp[-\nu_{\rm t}^2/2]$. Especially, the non-Gaussian signatures 
are pronounced at high absolute threshold such as $|\nu_{\rm t}|\gtrsim 1$ 
as a result of the strong coupling between the gravitational 
tidal shear and the large CMB curvature at such high threshold 
as explained.
For a Gaussian case, $F_Y(\nu_{\rm t})$ and $F_Z(\nu_{\rm t})$ 
should be equal to zero at all bins of $\nu_{\rm t}$ and thus
have both positive and negative large values of the cosmic variance
in each bin, although the mean values 
do not exactly converge to zero yet for the number of our realizations.
These results therefore mean that the non-Gaussian signatures induced by 
the lensing could be significantly distinguished compared to
the cosmic variance. 
In Figure \ref{fig:covmat} we show the contour map of 
$\skaco{(F_{Xi}-\skaco{F_{Xi}})(F_{Xj}-\skaco{F_{Xj}})}
/(\sigma_i\sigma_j)$ calculated from those realizations, 
where $\skaco{F_{Xi}}$ is 
the mean value and $\sigma_i$ the variance of the estimators
of $F_{X}(\nu_{\rm t})$ at threshold bin $\nu_i$. We have confirmed 
that the correlations for $F_Y(\nu_{\rm t})$ and $F_Z(\nu_{\rm t})$ 
are similar to the result in this figure. 
These correlations would
be required in order properly to quantify the significance of any
departure from Gaussian statistics when performing the fitting between
the numerical results and theoretical predictions. Figure 
\ref{fig:covmat} indicates and we have actually confirmed that, 
if we take the data at intervals of $\Delta\nu_{\rm t}\approx 0.5$, the
correlation matrix becomes to be very close to diagonal in each bin for 
all the cases we consider in this paper. 
Taking into account this result, the following results are shown in 
intervals of $\Delta\nu_{\rm t}=0.5$.

\begin{figure}[t]
 \begin{center}
     \leavevmode\epsfxsize=11cm \epsfbox{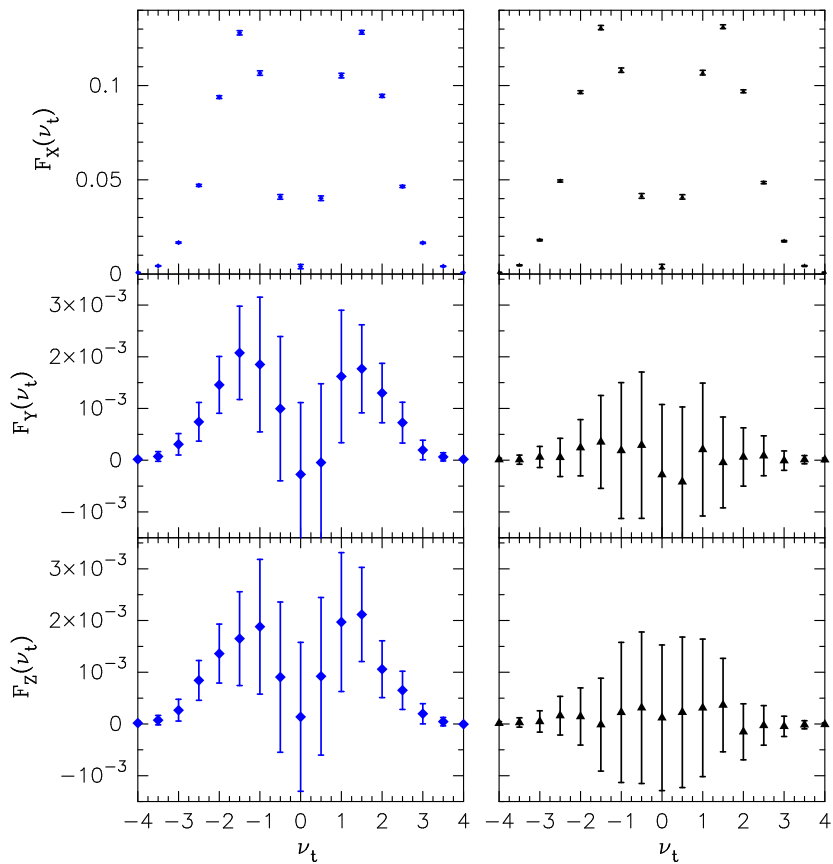}
\caption{
  Same as Figure \ref{fig:funglt3} with a 5.5 arcmin filtering
  scale and $\sigma_8=2.0$. All the plots are shown in intervals of
  $\Delta\nu_{\rm t}=0.5$ taking into account the result of Figure 
  \ref{fig:covmat}.
 \label{fig:funglt55}}
 \end{center}
\end{figure}

\begin{figure}[t]
 \begin{center}
     \leavevmode\epsfxsize=11cm \epsfbox{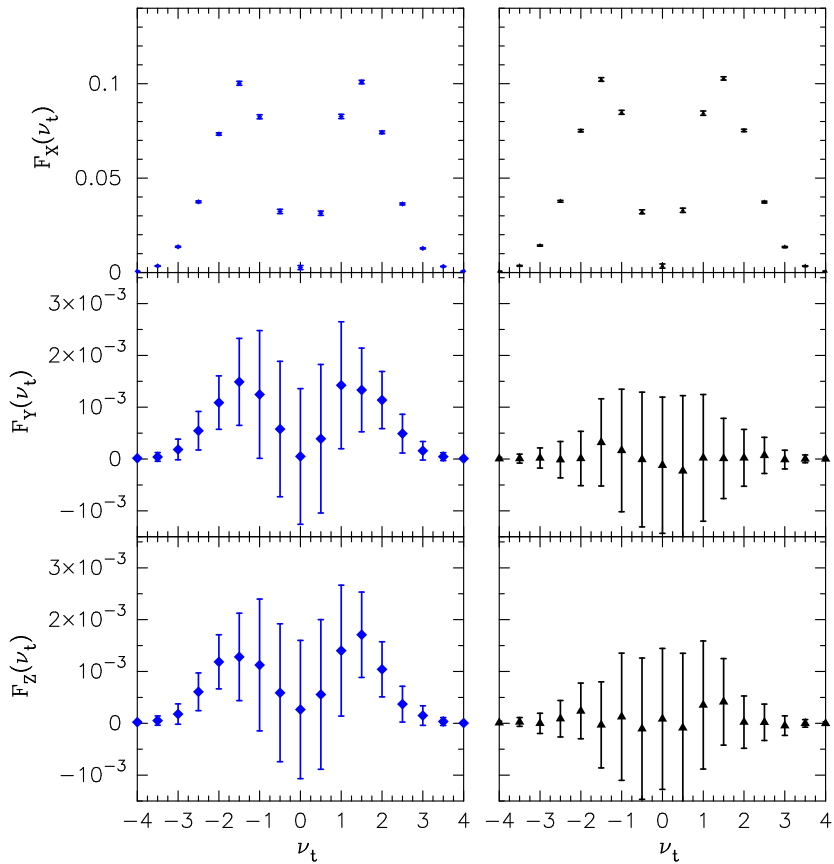}
\caption{
  This figure also includes the detector noise effect for the 
  case shown in Figure \ref{fig:funglt55},
  where we have assumed the noise level of Planck expressed in terms of 
  the variance of noise field per a $5.5$ arcmin FWHM pixel as 
  $\sigma_{\rm pix}=4.3\times 10^{-6}$. Note that the scales of $x$- and
  $y$- axes are same as in Figure \ref{fig:funglt55}. 
 \label{fig:funglt55wN}}
 \end{center}
\end{figure}

\begin{figure}[t]
 \begin{center}
     \leavevmode\epsfxsize=11cm \epsfbox{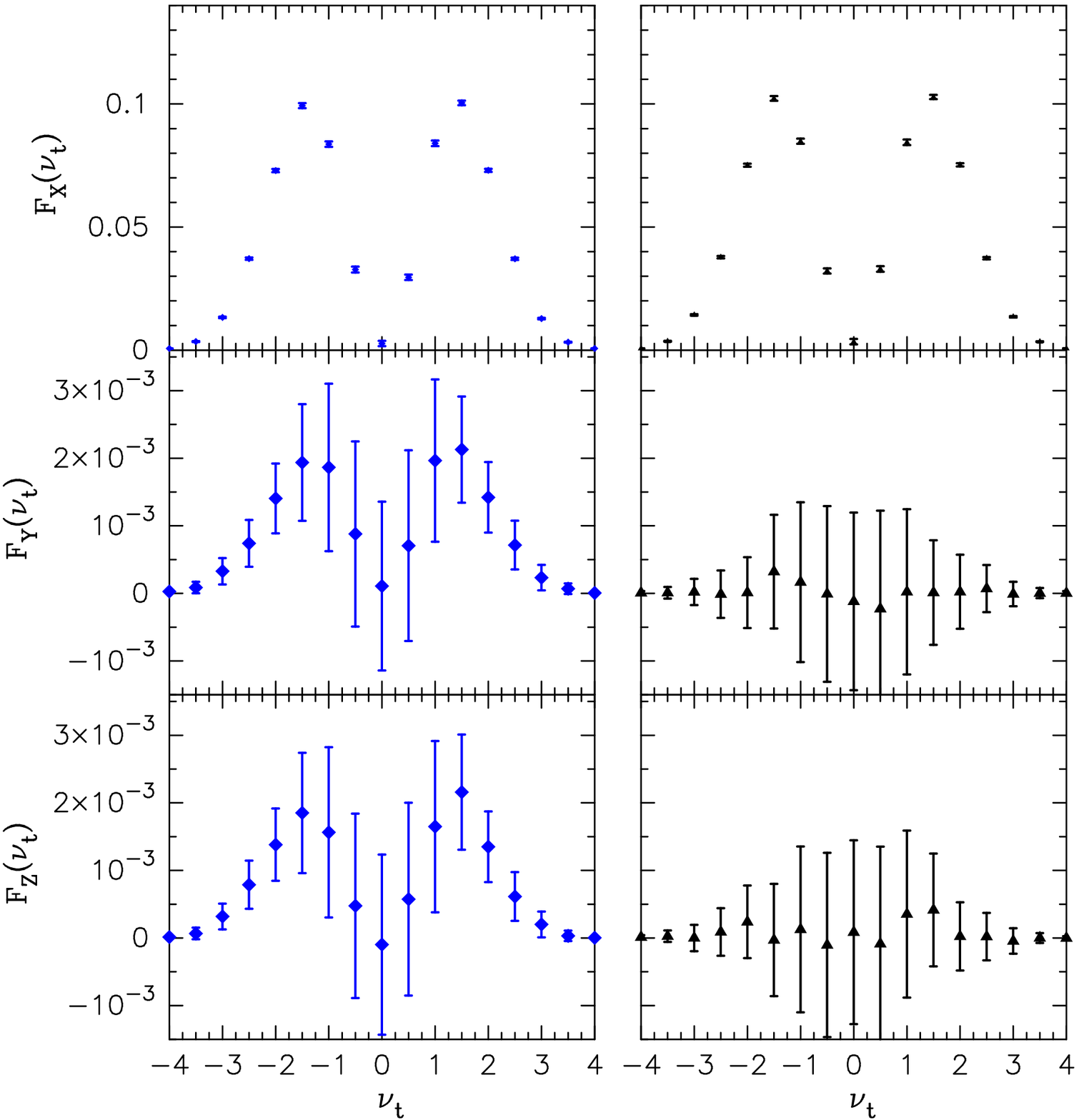}
\caption{
 Same as Figure \ref{fig:funglt55wN} with $\sigma_8=2.5$.
 \label{fig:funglt55s25}}
 \end{center}
\end{figure}

Figure \ref{fig:funglt55} demonstrates the results with the 
filtering scale of $5.5$ arcmin computed similarly as in Figure 
\ref{fig:funglt3}. One can see that increasing 
the filtering scale 
decreases the magnitude of lensing signals for $F_Y(\nu_{\rm t})$ and
$F_Z(\nu_{\rm t})$, 
because the filtering again tends to circularize the deformed
structures in the CMB map as pointed out by Bernardeau (1998).
However, even in this case of $\theta_{\rm fwhm}=5.5'$ 
the lensed curves of $F_Y(\nu_{\rm t})$ and $F_Z(\nu_{\rm t})$ 
remain having the distinct
functional dependence of $\nu_{\rm t}$ compared to the Gaussian case of
$F_Y(\nu_{\rm t})=F_Z(\nu_{\rm t})=0$.  In practice it is important  
to also take into account the detector noise effect on our
method. We here assume the instrumental specification of $217\rm{GHz}$
channel of the satellite mission {\em Planck Surveyor}~;
the noise level of $\sigma_N=4.3\times 10^{-6}$ per a pixel on a side of
the FWHM extent ($\theta_{\rm fwhm}=5.5'$). The noise
field at original fine grid is also convolved with the Gaussian filter of
FWHM scale to avoid domination of noise spikes at small angular scales 
(\cite{Barreiro98}). Figure \ref{fig:funglt55wN} shows the results.  
The noise effect reduces the amplitudes of $F_X(\nu_{\rm t})$ compared 
to that in Figure \ref{fig:funglt55} as a result of the change of quantity 
$\gamma_\ast(\equiv \sigma_1^2/\sigma_0\sigma_2)$ 
for the intrisic CMB anisotropies due to the noise. 
Even in this case, the figure clearly shows that the noise
level of Planck does not largely change the lensed shapes of $F_Y$ and $F_Z$
compared to the results of Figure \ref{fig:funglt55}, although the 
lensing signals at low thresholds such as $|\nu_{\rm t}|\simlt 1.5$ are 
weakened. Importantly, significant 
non-Gaussian signatures on $F_Y(\nu_{\rm t})$ 
and $F_Z(\nu_{\rm t})$ still remain compared to $F_Y(\nu_{\rm
t})=F_Z(\nu_{\rm t})=0$. 
This is because our method has so far relied on the normalized 
observable quantities such as $\nu(\bm{\theta})
=\Delta(\bm{\theta})/\sigma_0$ and those quantities are more 
robust against the systematic contributions of the detector
noise than the CMB fields themselves that are certainly affected by 
the noise. 
Figure \ref{fig:funglt55s25} shows
the results with $\sigma_8=2.5$ similarly as Figure \ref{fig:funglt55wN}. 
This figure explains that the lensing signals can significantly deviate from
the unlensed case if the lensing effect is adequately large. 


\subsection{Cosmological implications from the lensing-induced
  non-Gaussian signatures}

Another important question we should address is how useful cosmological
information on the large-scale structure formation 
can be extracted from the lensing signals onto the isotemperature 
statistics presented in the previous subsection. Since 
equation (\ref{eqn:glfun}) shows that the non-Gaussian
signatures depend on $\gamma_\ast$ and $\sigma_\kappa$,  
we become to consider the problem how
accurately the contribution of $\sigma_\kappa$ can be determined. 
As shown by Jain \& Seljak (1997), the magnitude of $\sigma_\kappa$
is sensitive to $\sigma_8$ and $\Omega_{\rm m0}$ parameters for the CDM 
models and also depends
on the filtering scale $\theta_{\rm s}$ for a realistic case. 
From equation (\ref{eqn:smconv}) we find that 
$\sigma_\kappa$ thus has following approximate scaling relations for 
flat universe models around
the fiducial model ($\Omega_{\rm m0}=0.3$, $\Omega_{\lambda0}=0.7$
 and $h=0.7$):
\begin{eqnarray}
&&\sigma_\kappa[3']\approx0.157\times(\theta_{\rm fwhm}/3')^{-0.35}
 \times(\sigma_8/2.0)^{1.1}\times
 (\Omega_{\rm m0}/0.3)^{0.25},
 \nonumber\\
&&\sigma_\kappa[10']
 \approx 9.67\times10^{-2}\times(\theta_{\rm fwhm}/10')^{-0.47}
 \times(\sigma_8/2.0)^{1.0}\times(\Omega_{\rm m0}/0.3)^{0.10}. 
\end{eqnarray}
Furthermore, since more fundamental information is contained in the
three-dimensional mass fluctuations, we have to take into account
the projection effect (\cite{JS}; \cite{SZPRL}; \cite{ZS99}; \cite{TF}). 
The convergence field at $\theta\simlt 10'$ depends mainly on
the mass fluctuations with wavelength modes of $\lambda\simlt 10
h^{-1}{\rm Mpc}$ and the structures distributed in wide redshift
ranges of $0\simlt z\simlt 10$ peaked at $z\approx 3$.  The lensing
distortion effect on the CMB map can thus  be a powerful
probe of the large-scale structures up to high redshift in principle, 
which is not attainable by any other means. 

\begin{figure}[t]
 \begin{center}
     \leavevmode\epsfxsize=10cm \epsfbox{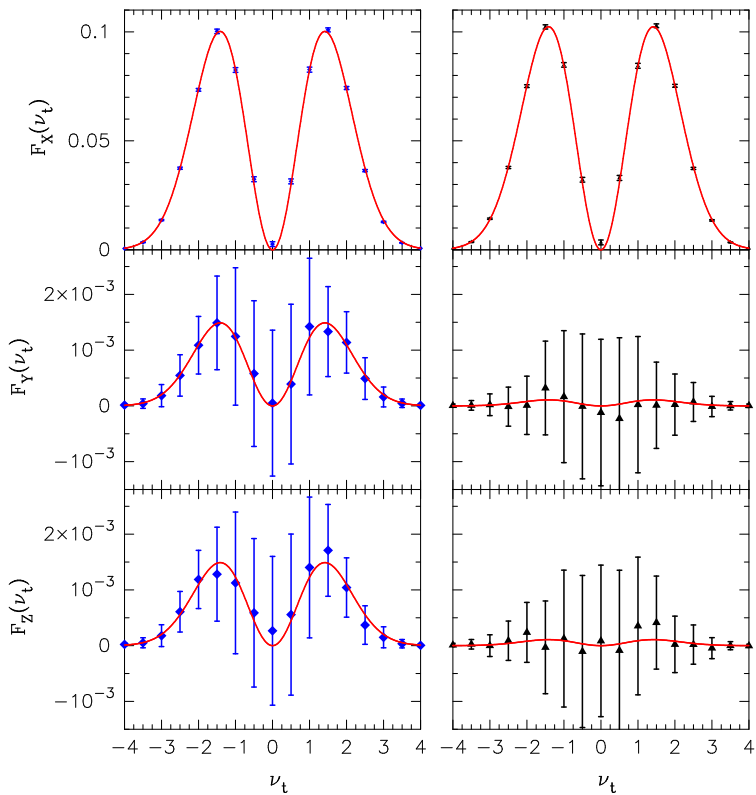}
\caption{
 The results of best-fit between the theoretical 
 predictions and the numerical results for the lensed (left: 
 $\sigma_8=2.0$) and unlensed (right) cases with 
 $\theta_{\rm fwhm}=5.5$ arcmin and the noise effect. 
 The line in each panel shows the best-fit theoretical 
 prediction (\ref{eqn:glfun}) (see Table \ref{tab:fit}).
\label{fig:bestfit}}
\end{center}
\end{figure}

Table \ref{tab:fit} summarizes the results obtained for
the determination of $\sigma_\kappa$ with a best-fit and the $1\sigma$
error, which arises from the cosmic variance or also
including errors due to the detector noise effect. 
The `analytic' value of $\sigma_\kappa$ in this table is calculated 
using the approximation for the 
beam smearing given by equation (\ref{eqn:smconv}). 
We here performed the 
$\chi^2$-fitting between the numerical results and 
theoretical predictions for $F_X(\nu_{\rm t})$, $F_Y(\nu_{\rm t})$ and 
$F_Z(\nu_{\rm t})$. Note that we have used each data of the functions 
in the range of $-4\le\nu_{\rm t}\le 4$ at intervals of 
$\Delta\nu_{\rm t}=0.5$, because the correlation matrix 
for those data is close to diagonal as explained in Figure \ref{fig:covmat}. 
Then, the theoretical predictions of $F_X(\nu_{\rm t})$, $F_Y(\nu_{\rm
t})$ and $F_Z(\nu_{\rm t})$ are given by two free parameters of 
$\gamma_\ast$ and $\sigma_\kappa$ expressed by 
equation (\ref{eqn:glfun}). Most importantly, 
Table \ref{tab:fit} clearly shows that 
$\sigma_\kappa$ estimated from the non-Gaussian signatures on $F_X$,
$F_Y$ and $F_Z$ could be significantly detected with high signal to
noise ratios compared to the unlensed case.
Here, $1\sigma$ error
for the $\sigma_\kappa$ determination corresponds to $\Delta\chi^2=2.3$
for the $\chi^2$ fitting.   
Figure \ref{fig:bestfit} demonstrates an example of the best-fit 
results for the noise case with $\sigma_8=2.0$ and 
$\theta_{\rm fwhm}=5.5$ arcmin. This figure shows that
$\sigma_\kappa$ is constrained mainly from the numerical data of
$F_Y(\nu_{\rm t})$ and $F_Z(\nu_{\rm t})$ at $|\nu_{\rm t}|\simgt 2$
that have relatively small cosmic variances. 
On the other hand, $\gamma_\ast$ is well constrained only by the data of
$F_X(\nu_{\rm t})$. Even if we use the value of $\tilde{\gamma}_\ast$ 
directly measured from the lensed simulated CMB maps instead of the 
fitting parameter $\gamma_\ast$ 
(see the paragraph under equation (\ref{eqn:glfun})), 
it causes only the slight change of results in Table \ref{tab:fit}.
One might then consider a  possibility to determine $\sigma_\kappa$ 
by comparing the measured $\tilde{\gamma}_\ast$ from the CMB maps 
to that of $\gamma_\ast$ obtained from the fitting of $F_X(\nu_{\rm t})$
through the relation of
$\tilde{\gamma}_\ast=\gamma_\ast(1-7\sigma_\kappa^2/2)$, but the
constraint is much weaker than that of using the non-Gaussian 
signatures on $F_Y(\nu_{\rm t})$ and $F_Z(\nu_{\rm t})$. 
Table \ref{tab:fit} also shows that the beam smearing effect is
crucial for our 
method and, in the case with $\theta_{\rm fwhm}=8'$ and $\sigma_8=2.0$
the lensing signatures are obscured by the cosmic variance. 
On the other hand, the detector noise effect does not largely 
affect the results. 
However, we have to note that the best-fit value of $\sigma_\kappa$ 
in all the considered cases is  underestimated compared to 
the analytic value of $\sigma_\kappa$ calculated 
by the approximation (\ref{eqn:smconv}) for beam smearing effect.
The possible reasons for the 
underestimation are ascribed to the effect of the discrete 
pixel in the simulated maps or to the mode coupling between the 
CMB field and lensing field caused by the filtering as explained 
in \S \ref{filter}. So far it is concluded based on the following results
that the reason is mainly due to the discrete pixel effect. 
We have confirmed that the best-fit 
value of $\gamma_\ast$ even for the unlensed case by our method  
also underestimates  the value of $\gamma_\ast$ 
calculated by the conventionally 
used approximation (e.g., \cite{BE}) expressed in terms of $C_l$ and 
$\theta_{\rm fwhm}$ as $\gamma_\ast\equiv 
\sigma_1^2(\theta_{\rm fwhm})/\sigma_0(\theta_{\rm
fwhm})\sigma_1(\theta_{\rm fwhm})$, where $\sigma_{\rm i}^2(\theta_{\rm
fwhm}) \equiv \int (ldl/2\pi)l^{2{\rm i}}C_l \exp[-l^2\theta_s^2]$ 
in the context of the small angle approximation and
$\theta_s=\theta_{\rm fwhm}/\sqrt{8\ln 2}$.  
For example, values of the best-fit and analytical prediction for 
$\gamma_\ast$ are $0.368$ and $0.38$, respectively, for unlensed case 
with $\theta_{\rm fwhm}=5.5$ and without the detector noise effect. 
For the discrete pixel data of simulated CMB maps, 
it is generally difficult to perform accurate measurements of 
statistical quantities defined from any {\em derivative} fields of CMB 
compared with their analytical predictions.  
Schmalzing, Takada \& Futamase (2000) have also
confirmed that it is crucial for accurate measurements of 
Minkowski functionals in a realistic CMB map 
to take into account the effect of discrete pixel, 
where we have used the interpolation technique.
Therefore, we have to further investigate the problem 
how the $\sigma_\kappa$ determination 
from a realistic CMB map performed by our method 
can reproduce its simple analytic prediction, 
for example, by using the numerical simulations combined to 
the interpolation technique. 

\begin{deluxetable}{lccc}
\tablewidth{0pt}
\tablecaption{
 Values of $\sigma_\kappa$ from best-fit
 parameterizations. Errors give the cosmic variance (or also includes
 the instrumental errors caused by detector noise) 
for $70\%$ sky coverage survey of the CMB map. Here, the analytic 
value of $\sigma_\kappa$ is calculated by the approximation 
(\ref{eqn:smconv}) for the beam smearing effect.}
\tablehead{$\sigma_8$
& filter scale & analytic $\sigma_\kappa^2\times 10^2$&
 $\sigma_\kappa^2\times10^2$, best-fit }
\startdata
$0.0$ (no lens)       &  $3'$          &-&   $0.02\pm0.03$ \\
$0.0$ (no lens)     &  $5.5'$          &-&   $0.04\pm0.03$ \\
$0.0$ (no lens)     &  $5.5'+$noise          &-&   $0.03\pm0.04$ \\
$1.5$       &  $3'$          &$1.31$&   $0.71\pm0.04$ \\
$1.5$       &  $5.5'$        &$0.86$&   $0.26\pm0.03$ \\
$1.5$       &  $5.5'+$noise  &$0.86$&   $0.23\pm0.04$ \\
$2.0$       &  $3'$          &$2.47$& $1.14\pm0.04$\\ 
$2.0$       &  $5.5'$        &$1.57$& $0.42\pm0.04$\\
$2.0$       &  $5.5'+$noise  &$1.57$& $0.41\pm0.05$\\
$2.0$       &  $8'$          &$1.15$& $0.07\pm0.04$\\
$2.5$       &  $5.5'$  &$2.51$&$0.62\pm0.04$\\
$2.5$       &  $5.5'+$noise  &$2.51$&$0.56\pm0.05$\\
 \enddata
 \label{tab:fit}
\end{deluxetable}

\section{Discussion and Conclusion}
In this paper, we developed a new simple method for extracting the
lensing-induced non-Gaussian signatures on the isotemperature
statistics in the CMB sky and also investigated the feasibility of
the method using the numerical experiments.
Importantly, by focusing on the characteristic three independent
functions obtained from the averages of quadratic combinations of
the second derivatives of CMB field over isotemperature contours with
each temperature threshold, it was found that the weak lensing
generates non-Gaussian signatures on those functions that have  
a distinct functional dependence of the threshold.
The result is a consequence of the coupling between the gravitational
tidal shear and the CMB curvature (defined by $-[\Delta_{,11}
+\Delta_{,22}]/\sigma_2$) through the intrinsic 
correlation between the CMB curvature and the temperature 
threshold predicted by the Gaussian theory. 
By means of the non-Gaussian
signatures,  it can be expected that the lensing signals are extracted
from an observed CMB data irrespective of the $C_l$ measurements 
or equivalently the assumptions for the fundamental cosmological 
parameters. Our numerical experiments indeed 
showed that the method allows us to extract the lensing signals with 
a high signal to noise ratio, provided that we have  
CMB maps with sufficiently low noise and high angular resolution as 
given by the Planck mission.
Recently, Seljak \& Zaldarriaga (1999) (see also \cite{ZS99}) developed
a powerful method for a direct reconstruction of the projected power
spectrum of matter fluctuations
 from the lensing deformation effect on the CMB maps. 
The method focuses on the averages of quadratic combinations of the 
gradient fields of CMB over a lot of
independent patches in the CMB sky like the analysis of the measurements
of the coherent gravitational distortion on images of distant galaxies
due to the large-scale structure. In the method, the lensing 
signatures on individual patches are extracted as 
differences between the measured statistical measures and 
their all-sky averages.
They showed that the reconstruction of input projected matter power 
spectrum could be successfully achieved if there is no 
sufficient small scale power of 
intrinsic CMB anisotropies. In this sense, however, 
the method partly depends on 
the statistical measurements of  intrinsic CMB anisotropies, and this is 
main difference between their and our method which we would like to 
stress. Moreover, our method focuses on the second derivative fields 
of CMB and, therefore, is more sensitive to the amplitudes of 
the projected matter power spectrum on smaller angular scales. 
Anyway, since the lensing signals in  the observed CMB sky
are weak,  we think that some independent statistical methods 
should be performed to extract them, which could also 
lead to constraints on the projected matter power spectrum 
at respective, different angular scales.  

Extending our method presented in this paper, 
one might consider a following possibility to extract the lensing 
distortion effect. 
BBKS and BE have shown that  structures around local higher maxima
or lower minima of temperature fluctuation field tend to have more peaked shape
and be more spherically symmetric around the peaks in the Gaussian
(unlensed) case.
From these features, it can be also expected that the weak lensing causes
stronger distortion effect on structures around the higher maxima (or 
lower minima) and
it might provide us more significant non-Gaussian signatures as a function
of the temperature threshold of peaks than our method did.
However, as quantitatively shown by Bernardeau
(1998), the statistical measure for the peaks provides the same or 
lower signal to noise ratio as or than the statistics for a field 
point taking into
account the cosmic variance, where he investigated the lensing effect 
on the probability distribution
 function of ellipticities around field point
or peak. This is because the number density of peaks in the CMB sky 
is not so sufficient  for these statistical measurements. For this reason, 
we  prospect a similar conclusion for the signal to noise ratio
obtained from the measurements of the lensing distortion effect on 
the structures around peaks using our method,
although this work will have to be further investigated carefully.

Recently, Schamlzing, Takada \& Futamase (2000) have shown that 
the lensing effect on the Minkowski functionals (the morphological
descriptions of the CMB map) appears just as a change 
of their normalization factors against the Gaussian predictions using 
the numerical experiments.  We indeed have confirmed that
the analytical predictions for the 
lensed Minkowski functionals done in the similar way as 
presented in this paper give the same conclusion as
the numerical results.  The result comes from the fact that 
 the lensing does not largely change the global topology
of the CMB anisotropies in a statistical sense. Likewise,
it is known that the gravitational potential from which
the shear is generated is invariant under the parity transformation and
the lensing does not induce the so-called `B-type' polarization defined
from combinations of the derivative fields of CMB 
fluctuations (\cite{SZPRL}; \cite{ZS99}).
These results mean that the weak lensing cannot simply generate a new 
mode of pattern of the CMB anisotropies that is absent in the Gaussian
case. For these reasons, in this paper we focus on 
the another information on
statistical properties of the intrinsic CMB that is useful for the study
of lensing effect and can be specifically predicted 
based on the Gaussian  random theory. 
Another issue we should discuss is the possible effect on our results 
caused by non-Gaussian features on the convergence field 
of the large-scale structure at $\theta<10'$ that are  revealed by
the ray-tracing simulations (\cite{JSW}; \cite{Hamana}), which we have
ignored in the numerical experiments of the lensed CMB maps. 
However, since 
the lensing effect on the CMB can be treated as a mapping, which is 
 expressed as 
$\tilde{\Delta}(\bm{\theta})=\Delta(\bm{\theta}+\bm{\xi})$,  and
the lensing contributions to the CMB are always coupled to the
contributions from the primary 
CMB fields, we prospect that the effect will not change the 
functional dependence of temperature threshold on the lensing-induced 
non-Gaussian signatures of $F_X(\nu_{\rm t})$, $F_Y(\nu_{\rm t})$ 
and $F_Z(\nu_{\rm t})$ expressed by 
equation (\ref{eqn:glfun}), even if the effect could enhance the magnitudes. 

Undoubtedly, other secondary effects could induce non-Gaussian properties 
in the observed CMB map. The most important
effects are the (thermal) Sunyaev-Zel'dovich effect and
the foreground contaminations such as Galactic foreground and extragalactic
point sources. Those effects can be removed to some extent by using advantages
of their spectral properties, although further reliable investigations
should be done for any measurements of CMB.
Furthermore, in this paper we have presented the lensing-induced 
non-Gaussian signatures on {\em three}
independent functions, $F_X(\nu_{\rm t}),$ $F_Y(\nu_{\rm t})$ 
and $F_Z(\nu_{\rm t})$, and therefore 
we expect that the property of $F_Y(\nu_{\rm t})=F_Z(\nu_{\rm t})$ 
can provide us
a clue to resolving the lensing contributions from the other
secondary effects.

\section*{Acknowledgments}
The author would like to thank T. Futamase and J. Schmalzing for
fruitful discussions and  valuable comments.  
He also acknowledges the useful comments from anonymous referee. 
He thank 
U. Seljak and M. Zaldarriaga for making  their CMBFAST code publicly
available. He also acknowledges a support from a Japan Society
for Promotion of Science (JSPS) Research Fellowship.

\end{document}